\documentclass{mem}
\usepackage{natbib}\usepackage{txfonts}\usepackage{balance}
\usepackage{graphicx}
\idline{00}{000}
\begin{document}

\newcommand{\dsct}{$\delta$~Sct\,}

\title{
The overlooked role of stellar variability in LMC intermediate-age clusters
}

   \subtitle{}

\author{Ricardo Salinas}

\institute{
Gemini Observatory, Casilla 603, La Serena, Chile    
\email{rsalinas@gemini.edu}
}

\authorrunning{R. Salinas}

\titlerunning{Variability in intermediate-age LMC clusters}

\abstract{
Broad, extended main sequence turnoffs seen in the majority of the intermediate-age (1--3 Gyr) LMC star clusters, have been interpreted as the result of an extended star formation history and/or the effect of extreme stellar rotation. A more fundamental explanation may be given by stellar variability. For clusters in these age range, the instability strip crosses the upper main sequence producing a number of variable stars (known as Delta Scuti) which, if nor properly taken into account, could appear as an extended turnoff. First results of a variability program in the LMC cluster NGC 1846 reveals a sizeable number of this type of variables, although still too low to produce a meaningful broadening, with the caveat that the true variable content of the center of this and other clusters in the LMC will only be revealed with a dedicated \textit{HST} program.

}
\maketitle{}

\section{Introduction}

The large range of ages found for the star cluster population in the Magellanic Clouds give us the chance to study stellar evolutionary stages and environments not available in our Galaxy. 

One of the most striking observations is the extension of the main-sequence turn-offs (MSTOs) beyond the canonical expectation for single stellar populations that a large number of intermediate-age ($\sim$1--3 Gyr) star clusters in the Large Magellanic Cloud (LMC) present \citep[e.g.][]{mackey07,milone09}. This extension might be explained by an extended star formation history \citep[e.g.][]{goudfrooij09,milone09} or as a signature of stellar rotation \citep[e.g.][]{bastian09,brandt15}.

\section{The role of stellar variability}
\begin{figure*}[t!]
\resizebox{\hsize}{!}{\includegraphics{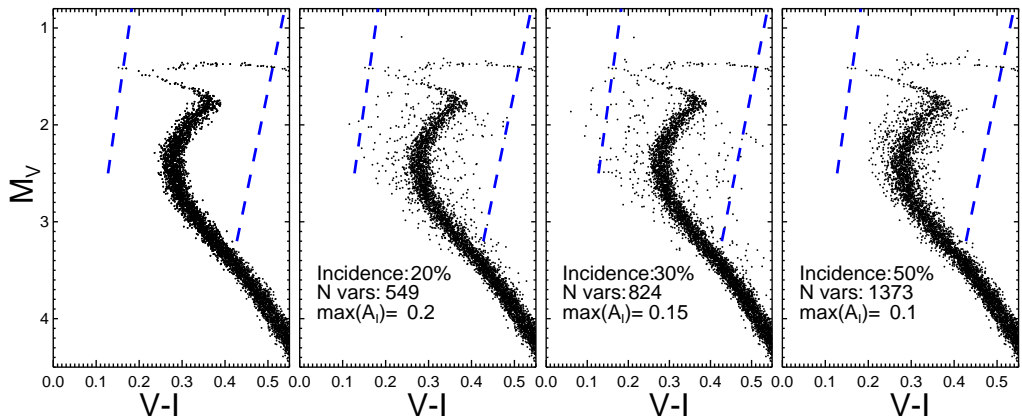}}
\caption{\footnotesize  A model of the influence of \dsct based on BaSTi synthetic CMDs \citep{pietrinferni04}. In each panel, the dashed blue lines indicate the instability strip. Each panel indicates the assumed incidence, the maximum amplitude for the light curve amplitude distribution and the total number of \dsct for each model. Figure adapted from \citet{salinas16b}. }
\label{Fig1}
\end{figure*}

Even though both scenarios have been shown to have advantages and shortcomings when explaining the morphology of the color-magnitude diagrams (CMDs) of LMC intermediate-age clusters \citep[e.g.][]{bastian16,goudfrooij17}, a more fundamental ingredient has been commonly overlooked: stellar variability. The lower instability strip crosses the upper main sequence and MSTO for stellar populations of ages between 1 and 3 Gyr, therefore producing pulsations in a number of stars. These pulsating main-sequence stars are known as Delta Scuti (hereatfer \dsct, see e.g. \citealt{breger00} for a review).

When images of a star cluster are then obtained only a couple of times per filter, enough to construct a CMD, \dsct will be observed at random phases, away from their ``static" colors and luminosities, and therefore introducing a spurious spread which could be mistaken for a genuine extended MSTO \citep{salinas16b}. 

This idea is exemplified in Fig. \ref{Fig1}. Starting from a synthetic CMD based on BaSTi models \citep{pietrinferni04} (leftmost panel), we select a percentage of stars within the instability strip (represented with dashed blue lines) which will be variables. This percentage is commonly known as the incidence of \dsct. For each selected star we model a \dsct light curve, which is then ``observed'' at a given fixed time. Details can be seen in \citet{salinas16b}. Fig. \ref{Fig1} shows how the MSTO is broadened as a function of incidence and the amplitude distribution of the light curves, without a significant broadening of either the upper MS or the sub-giant branch, in close resemblance to the effect observed in \textit{HST} CMDs.

\section{A test case: NGC 1846}

In order to the test the hypothesis of \citet{salinas16b}, we obtained time series photometry of the LMC cluster NGC 1846 using the Gemini South telescope equipped with GMOS used as an imager (Gemini program ID GS-2015B-FT-7). Sixty six 120 second frames were obtained with the SDSS $r$ filter, within the time span of 3.2 hours. After image reduction with the Gemini/IRAF package, variable stars were searched and its light variations were measured using a combination of ISIS \citep{alard00} and DAOPHOT \citep{stetson87}, following the procedure described in \citet{salinas16a}.

Our search revealed the presence of 55 \dsct in the NGC 1846 field covered by GMOS, plus 18 variables of other types \citep{salinas18}. Considering the effect of radial completeness and background contamination, and using the cluster structural parameters from \citet{goudfrooij09}, we estimate the total number of \dsct in the cluster between 45 and 60 members.

If we compare with the hundreds or even thousands that are necessary to produce a significant broadening of the MSTO according to the \citet{salinas16b} models, then it seems the contribution of \dsct to the extended MSTO phenomenon would be negligible, although in the Gemini data, given the extreme crowding, basically we found no \dsct within the half-light radius of the cluster and therefore an unreliable extrapolation is ensued.

\begin{figure}[t!]
\resizebox{\hsize}{!}{\includegraphics{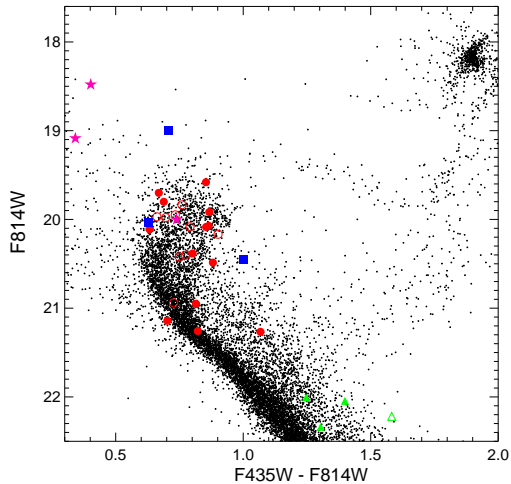}}
\caption{\footnotesize \textit{HST} photometry of NGC 1846 from \citet{milone09}. Filled symbols indicate variable stars for which no signs of variability were found by \citet{milone09}. Figure adapted from \citet{salinas18}.}
\label{Fig2}
\end{figure}

Even though the ground-based Gemini photometry is too coarse to reveal a broadening of the MSTO,  some insight on the effect of these variables on a CMD morphology can be done cross-matching the variable star catalogue with the \textit{HST} photometry of \citet{milone09}. Fig. \ref{Fig2} shows 33 variables found in the Gemini data that are within the ACS field of view. \dsct are depicted with red symbols, and as expected, they are concentrated at the upper MS and MSTO. The most revealing result from this plot comes from the quality flag assigned by \citet{milone09} to each star in their photometry, where one of the criteria is based on the rms of the magnitudes measured in different frames. Filled symbols in Fig. \ref{Fig2} are those stars labeled as ``good quality'', that is, 24 out 33 of these variable stars, despite a rejection based on rms, are considered with good quality. Since rms was not the only criterion used by \citet{milone09} to discard stars, it is possible to say that at least a 70\% of variables will not be found with current archival data of LMC clusters. Not even RR Lyrae stars, which have much larger amplitudes than \dsct, will be detected (blue filled symbols).

\section{Conclusions}

Extended MSTOs in intermediate-age clusters of the LMC may be the product of extended star formation, stellar rotation and/or stellar variability. A first test of the influence of variability in the morphology of the MSTO has been attempted using Gemini photometry of the LMC cluster NGC 1846, revealing a large but insufficient amount of variability at the MSTO level, albeit with great uncertainty on the true variable star content within its half light radius.

New observations scheduled with the Gemini South and SOAR telescopes will shed new light into the role of variability in lower mass LMC clusters, and for the first time, in SMC clusters.

\begin{acknowledgements}
RS thanks Kathy Vivas, Jay Strader, Michael Pajkos and Rodrigo Contreras-Ramos with whom this collaboration has been developed. Supported by the Gemini Observatory, which is operated by the Association of Universities for Research in Astronomy, Inc., on behalf of the international Gemini partnership of Argentina, Brazil, Canada, Chile, and the United States of America.
 
\end{acknowledgements}

\bibliographystyle{aa}
%\bibliography{salinas}

\begin{thebibliography}{14}
\expandafter\ifx\csname natexlab\endcsname\relax\def\natexlab#1{#1}\fi

\bibitem[{{Alard}(2000)}]{alard00}
{Alard}, C. 2000, \aaps, 144, 363

\bibitem[{{Bastian} \& {de Mink}(2009)}]{bastian09}
{Bastian}, N. \& {de Mink}, S.~E. 2009, \mnras, 398, L11

\bibitem[{{Bastian} {et~al.}(2016){Bastian}, {Niederhofer},
  {Kozhurina-Platais}, {Salaris}, {Larsen}, {Cabrera-Ziri}, {Cordero},
  {Ekstr{\"o}m}, {Geisler}, {Georgy}, {Hilker}, {Kacharov}, {Li}, {Mackey},
  {Mucciarelli}, \& {Platais}}]{bastian16}
{Bastian}, N., {Niederhofer}, F., {Kozhurina-Platais}, V., {et~al.} 2016,
  \mnras, 460, L20

\bibitem[{{Brandt} \& {Huang}(2015)}]{brandt15}
{Brandt}, T.~D. \& {Huang}, C.~X. 2015, \apj, 807, 25

\bibitem[{{Breger}(2000)}]{breger00}
{Breger}, M. 2000, in Astronomical Society of the Pacific Conference Series,
  Vol. 210, Delta Scuti and Related Stars, ed. M.~{Breger} \& M.~{Montgomery},
  3

\bibitem[{{Goudfrooij} {et~al.}(2017){Goudfrooij}, {Girardi}, \&
  {Correnti}}]{goudfrooij17}
{Goudfrooij}, P., {Girardi}, L., \& {Correnti}, M. 2017, \apj, 846, 22

\bibitem[{{Goudfrooij} {et~al.}(2009){Goudfrooij}, {Puzia},
  {Kozhurina-Platais}, \& {Chandar}}]{goudfrooij09}
{Goudfrooij}, P., {Puzia}, T.~H., {Kozhurina-Platais}, V., \& {Chandar}, R.
  2009, \aj, 137, 4988

\bibitem[{{Mackey} \& {Broby Nielsen}(2007)}]{mackey07}
{Mackey}, A.~D. \& {Broby Nielsen}, P. 2007, \mnras, 379, 151

\bibitem[{{Milone} {et~al.}(2009){Milone}, {Bedin}, {Piotto}, \&
  {Anderson}}]{milone09}
{Milone}, A.~P., {Bedin}, L.~R., {Piotto}, G., \& {Anderson}, J. 2009, \aap,
  497, 755

\bibitem[{{Pietrinferni} {et~al.}(2004){Pietrinferni}, {Cassisi}, {Salaris}, \&
  {Castelli}}]{pietrinferni04}
{Pietrinferni}, A., {Cassisi}, S., {Salaris}, M., \& {Castelli}, F. 2004, \apj,
  612, 168

\bibitem[{{Salinas} {et~al.}(2016{\natexlab{a}}){Salinas}, {Contreras Ramos},
  {Strader}, {Hakala}, {Catelan}, {Peacock}, \& {Simunovic}}]{salinas16a}
{Salinas}, R., {Contreras Ramos}, R., {Strader}, J., {et~al.}
  2016{\natexlab{a}}, \aj, 152, 55

\bibitem[{{Salinas} {et~al.}(2016{\natexlab{b}}){Salinas}, {Pajkos}, {Strader},
  {Vivas}, \& {Contreras Ramos}}]{salinas16b}
{Salinas}, R., {Pajkos}, M.~A., {Strader}, J., {Vivas}, A.~K., \& {Contreras
  Ramos}, R. 2016{\natexlab{b}}, \apjl, 832, L14

\bibitem[{{Salinas} {et~al.}(2018){Salinas}, {Pajkos}, {Vivas}, {Strader}, \&
  {Contreras Ramos}}]{salinas18}
{Salinas}, R., {Pajkos}, M.~A., {Vivas}, A.~K., {Strader}, J., \& {Contreras
  Ramos}, R. 2018, ArXiv e-prints

\bibitem[{{Stetson}(1987)}]{stetson87}
{Stetson}, P.~B. 1987, \pasp, 99, 191

\end{thebibliography}

\end{document}